\documentclass[prd,twocolumn,showpacs,preprintnumbers,amsmath,amssymb,nofootinbib]{revtex4}
\usepackage{graphicx}

\DeclareMathOperator{\arccosh}{arccosh}

\begin{document}
\title{Dissociation by acceleration}
\author{Kasper Peeters}
\email[Email:]{kasper.peeters@aei.mpg.de}
\affiliation{Institute for Theoretical Physics, Utrecht University, P.O.~Box 80.195,
3508 TD Utrecht, The Netherlands.}
\author{Marija Zamaklar}
\email[Email:]{marija.zamaklar@durham.ac.uk}
\affiliation{Department of Mathematical Sciences, Durham University,
  South Road, Durham DH1 3LE, United Kingdom.}

\preprint{ITP-UU-07/59, SPIN-07/45, DCPT-07/63}
\begin{abstract}
  We show that mesons, described using rotating relativistic strings
  in a holographic setup, undergo dissociation when their acceleration~$a$
  exceeds a value which scales with the angular momentum~$J$
  as~$a_{\text{max}} \sim \sqrt{T_s/J}$, where~$T_s$ is the string tension.
\end{abstract}

\maketitle

\section{Introduction and summary}

Rindler horizons provide an intriguing arena in which one can probe
aspects of event horizons using simple accelerated probes in flat
space-time. One such aspect concerns Fulling-Unruh radiation, which
makes accelerated probes feel a thermal bath. While the temperature of
this heat bath is relatively low in everyday situations, it has been
suggested that large enough accelerations may be achievable in
particle accelerators. The idea proposed
in~\cite{bell1b,Bell:1984sr,Bell:1986ir,Korsbakken:2004bv} is to use
polarised \emph{electrons} in a storage ring as thermometers. Their
depolarisation rate can then be used as a measure of the local
temperature.

In the present paper we want to analyse, instead, the thermal effects
of acceleration on \emph{mesons}. In order to describe mesons, we will
use generic ideas from holographic models for gauge theories with
matter.\footnote{The idea of exploring the thermal effect of
  acceleration on hadrons has previously been explored
  by~\cite{Castorina:2007eb}, with somewhat similar qualitative
  conclusions, but their methods are non-holographic in nature and
  rather different from the ones used in the present paper.}
Large-spin mesons are described in these models by long, rotating
strings~\cite{Kruczenski:2004me} which end on flavour
branes~\cite{Karch:2002sh}. They bend into the
holographic direction, forming a U-like shape. The constituent masses
of the quark and anti-quark map to the lengths of the string segments
in the holographic direction. In the limit in which the constituent
quark masses vanish, mesons are thus simply massless relativistic
strings rotating at the position of the ``infrared wall'' or
generalisations thereof. This massless limit is the one we will be
considering here.

If we now accelerate this rotating string in one of the three
space-like directions, a Rindler horizon forms. The situation is then
reminiscent of the analysis of rotating relativistic strings in
holographic setups at finite temperature. In those setups, it is known
that a horizon at some position in the \emph{holographic} direction leads to
the appearance of a maximum energy and spin, beyond which holographic
mesons melt~\cite{Peeters:2006iu}. By analogy with those results, we
expect that the acceleration of mesons in the three dimensional
space-like directions, and the related Rindler horizon which is now
located entirely in the \emph{four-dimensional} world, will also lead to a
dissociation effect.\footnote{Static Wilson-line configurations
  in Rindler space-time have been analysed
  in~\cite{Berenstein:2007tj}, but these do not exhibit all the
  phenomena which we find for rotating configurations. More
  importantly, the main difference of our analysis with respect
  to~\cite{Berenstein:2007tj} is the interpretation of the
  results. While~\cite{Berenstein:2007tj} discusses acceleration in
  the fifth, holographic direction and relates acceleration to the
  dissociation temperature in the dual gauge theory, the maximal
  acceleration discussed here is a genuine four-dimensional one.}

Our results indeed confirm this expectation. We find that spinning
strings have an upper bound for their acceleration, which is set by
the inverse square root of the angular momentum, \mbox{$a_{\text{max}}
  \sim\sqrt{T_s/J}$}.  As spin-off, we also discuss an
extension of this setup, in which the Rindler horizon is used to model
aspects of a holographic horizon, obtaining the velocity-dependence of
the dissociation length as
in~\cite{Peeters:2006iu,Liu:2006nn,Chernicoff:2006hi}.

\section{Rotating strings in Rindler space}

\begin{figure}[t]
\includegraphics[width=.7\columnwidth]{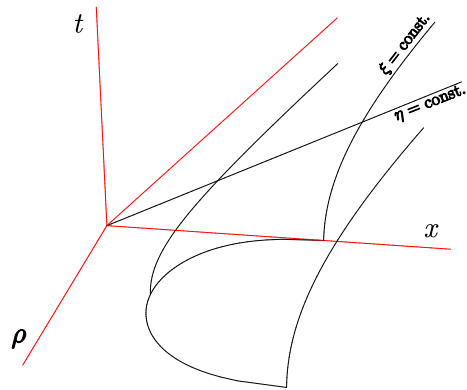}
\caption{The coordinate system used to describe rotating accelerated
  strings. The direction~$\phi$ in which the rotation takes place is suppressed.\label{f:accel_coords}}
\end{figure}

In order to describe accelerated spinning strings, we will use a
Rindler coordinate system. A particle which experiences a constant
acceleration in the~$x$ direction satisfies the equation of motion
\begin{equation}
\frac{{\rm d}}{{\rm d}t} \frac{m\, {\rm d}x/{\rm
    d}t}{\sqrt{\displaystyle 1-({\rm d}x/{\rm
        d}t)^2}} = F\,,
\end{equation}
with~$a = F/m$. Solutions to this equation are given by hyperbolas,
\begin{equation}
x^2 - t^2 = a^{-2}\, \quad \text{i.e.} \quad x =
\frac{1}{a}\cosh\left(a \eta\right)  ,
\, t = \frac{1}{a}\sinh\left(a \eta\right) .
\end{equation}
To describe accelerated motion, it is convenient to introduce
Rindler coordinates~$(\eta,\xi)$ which are adapted to these world-lines, in
the sense that fixed-$\xi$ curves map to world-lines of
constant acceleration $a=1/\xi$. More precisely, the coordinate
transformation is given by
\begin{equation}
x = \xi\,\cosh(\kappa \eta )\,,\quad
t = \xi\,\sinh(\kappa \eta)\,.
\end{equation}
The metric then takes the form 
\begin{equation}
\label{Rind}
{\rm d}s^2 = - \xi^2 \kappa^2 {\rm d}\eta^2 + {\rm d}\xi^2 + {\rm d}\rho^2 +
\rho^2\,{\rm d}\phi^2 \,,
\end{equation}
where we have added an additional two flat dimensions to accommodate
the rotating strings which we intend to study. The coordinate system
is once more summarised in figure~\ref{f:accel_coords}.

Our strings will be described by the following ansatz in terms of the
world-sheet coordinates~$(\tau,\sigma)$,
\begin{equation}
\label{ansz1}
\eta = \eta(\tau)\,,\quad \xi=\xi(\sigma)\,,\quad\rho=\rho(\sigma)\, , \quad \phi = \omega\tau\,.
\end{equation}
The Nambu-Goto action for our rotating string now reads
\begin{equation}
\label{e:action}
S = T_s \int\!{\rm d}\tau\int_{-L/2}^{L/2}\!{\rm d}\sigma\,
\sqrt{ \big( \xi^2 \kappa^2 {\dot \eta}^2 - \omega^2\rho^2 \big) \big(\rho'^2 + \xi'^2\big)}\,.
\end{equation}
We will soon make the gauge choice~$\eta = \tau$ which simplifies this
action further.  The symbol $T_s$ denotes the string tension of QCD
string.  In the holographic framework, this tension is related to the
tension of the fundamental string via the warping factor in the metric
in the radial (holographic) direction. Our string will not move
freely, following the action~\eqref{e:action}, as we will also add an
external force which acts on the endpoints and accelerates the
string. The effect of the force will be described by imposing
appropriate boundary conditions in the $\xi$-direction, as we will
explain in more detail below.

The ansatz~\eqref{ansz1} describes a string which accelerates in the
direction \emph{orthogonal} to the rotation plane. It is of course
also possible to accelerate the meson in a direction which is under an
arbitrary angle with respect to this plane, or even in the plane of
rotation. Describing such a string configuration is far more involved
as the component of angular momentum orthogonal to the acceleration
is not conserved. However, we will argue at the end of the next
section that acceleration in this case also leads to dissociation of
the meson.\footnote{Note that such an acceleration cannot be interpreted
  as gravitational (i.e.~in the spirit of~\cite{Berenstein:2007tj}).}

There are two conserved charges which can be used to characterise the
string. First, there is the angular momentum $J$, associated to the
rotation in $\phi$ direction, and second there is the boost charge, associated
to the translations in the $\eta$-direction.  The energy, (associated
with translations in the $t$-direction) is not a conserved
quantity, as the boundary conditions applied to the string endpoints break
this symmetry, and lead to a constant increase of the string energy.
Explicitly, the angular momentum and boost are given by
\begin{equation}
\begin{aligned}
J & = T_s \int_{-L/2}^{J/2}\!{\rm d}\sigma\,
   \frac{\rho^2\,\omega}{X(\sigma)} \, , \\[1ex]
B &=  T_s \int_{-L/2}^{J/2}\!{\rm d}\sigma\, \frac{ (- \kappa^2 \xi^2)\big(1 +
  (\xi')^2\big)}{X(\sigma)} \, , 
\end{aligned}
\end{equation}
where~$X(\sigma)$ is defined as
\begin{equation}
X(\sigma) \equiv\sqrt{ \frac{\xi^2 \kappa^2  - \rho^2\omega^2}{\rho'^2 + \xi'^2}}\,.
\end{equation}


\section{Accelerated mesons}

\subsection{Solution for the accelerated meson}

The \emph{bulk} equations of motion (i.e.~ignoring the surface terms)
for the fields $\rho$ and $\xi$ which follow from
the  action~\eqref{e:action} read
\begin{subequations}
\begin{align}
\label{e:eomrho}
-\frac{{\rm d}}{{\rm d}\sigma}\Big( \rho' X(\sigma)\Big) -
\frac{\rho\omega^2}{X(\sigma)} &= 0\,,\\[1ex]
\label{e:eomxi}
- \frac{{\rm d}}{{\rm d}\sigma} \Big( \xi' X(\sigma)\Big) 
+ \frac{\xi \kappa^2}{X(\sigma)} &= 0\,.
\end{align}
\end{subequations}
It is easy to check that equations of motion for the other fields
($\eta$ and $\phi$) are automatically satisfied with the ansatz (\ref{ansz1}).

At the boundaries of the string there are additional terms which need
to be taken into account. The external force acts only in the
$\xi$-direction, while there are no forces in the other directions;
hence the surface terms for all fields except~$\xi$ have to vanish.
The surface terms for the fields $\rho,\phi$ and $\eta$ are all
proportional to $X(\sigma)$, which allows for two possible boundary conditions,
\begin{subequations}
\begin{equation}
\label{st1}
\xi'|_{\sigma = \pm L/2} = \infty  \,,
\end{equation}
or
\begin{equation}
\label{st2}
 \rho|_{\sigma = \pm L/2} =
\frac{\kappa}{\omega}\xi(\pm L/2)\,.
\end{equation}
\end{subequations}
The surface term for the field $\xi$ is
\begin{eqnarray}
\label{xi-bc}
\left.\frac{\delta S}{\delta \xi'} \delta \xi \right|_{\sigma = \pm L/2} = \xi'
X(\sigma)  \delta \xi  \bigg{|}_{\sigma = \pm L/2}\, .
\end{eqnarray}
We see that if we were to impose the second boundary
condition~\eqref{st2}, the boundary term~\eqref{xi-bc} for~$\xi$ would
vanish automatically, which clearly does not describe an accelerated
string with a force applied to its endpoints. Hence, we choose the
condition ~$\xi'=\infty$ which is essentially also the boundary
condition imposed on the spinning string solutions
of~\cite{Kruczenski:2004me}. For this boundary condition, the product
of the first two factors in~\eqref{xi-bc} does not vanish. By adding
an external force, we impose the Dirichlet boundary
condition~$\delta\xi=0$, which forces the endpoints to move with
constant acceleration~$\xi^{-1}$.

After choosing the gauge~$\rho=\sigma$, the equation of
motion~\eqref{e:eomrho} can now be integrated once to give
\begin{equation}
X(\sigma)^2 = - \omega^2\rho^2 + C^2\,,
\end{equation}
where~$C$ is an integration constant. Inserting the definition
of~$X(\sigma)$ we obtain the differential equation
\begin{equation}
\xi'^2 = \frac{\xi^2 \kappa^2  - C^2}{C^2 - \omega^2\sigma^2}\,.
\end{equation}
This can be integrated to give the solution
\begin{equation}
\xi(\sigma) = \frac{C}{\kappa} \cosh \left[ \frac{\kappa}{\omega} \arcsin\left(
  \frac{\omega\sigma}{C}\right) + D \right]\,,
\end{equation}
where~$D$ is the second integration constant which will set to zero
from now on (without any loss of generality, as this just means that
we choose the coordinate origin in the $\sigma$~direction in a
symmetric way, i.e.~such that \mbox{$\sigma=0$} corresponds to the tip
of the U-shaped string). This solution also satisfies~\eqref{e:eomxi},
and in the limit~$\omega\rightarrow 0$ it reduces to the solution
found in~\cite{Berenstein:2007tj}.

\begin{figure}[t]
\begin{center}
\includegraphics[width=.7\columnwidth]{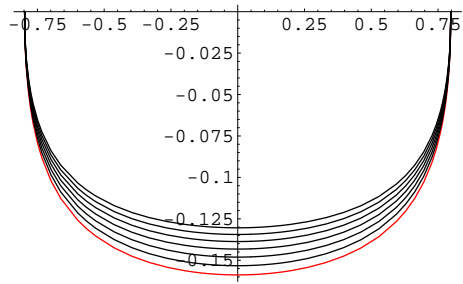}
\end{center}
\caption{The shape of the string in the $(x,\rho)$ plane as it evolves in
  time, in an accelerated frame where the endpoint positions are fixed. 
  The red (bottom) curve represents the shape at~\mbox{$t=0$}. \label{f:shape-string-xt}}
\end{figure}

We now impose the boundary condition, i.e.~that \mbox{$\xi'(\pm L/2)
  \rightarrow \pm \infty$}, which fixes the length of the string in
terms of the angular frequency and the constant~$C$,
\begin{equation}
\frac{L}{2} = \frac{C}{\omega}\,.
\end{equation}
Using this relation to eliminate~$\omega$ from the solution, we find
\begin{equation}
\label{e:xiLC}
\xi(\sigma) = \frac{C}{\kappa} \cosh\left[ \frac{\kappa L}{2 C} \arcsin\left(
  \frac{2\sigma}{L} \right) \right]\,.
\end{equation}
The total angular momentum for this solution reduces, after
elimination of~$\omega$, to the simple form
\begin{equation}
\label{JL}
J = \frac{\pi}{8} T_s L^2\, ,
\end{equation}
while the boost charge does not seem to have a simple analytic
expression.
  
In terms of the angular momentum, the shape of the string in the
($\xi,\rho=\sigma$) plane is
\begin{equation}
\label{e:stringshape}
\xi(\sigma) = \frac{C}{\kappa} \cosh\left[ \frac{\kappa}{C} \sqrt{\frac{2 J}{\pi T_s}} \arcsin \left( \sqrt{\frac{\pi T_s}{2 J}} \sigma \right) \right]\,.
\end{equation}
It is more intuitive to describe the shape of the string in the
 original Minkowski coordinates $(x,\rho)$ 
\begin{equation}
\label{e:shapext}
x(t,\rho) = \sqrt{ \frac{C^2}{\kappa^2} \cosh^2\left[ \frac{\kappa}{C} \sqrt{\frac{2
      J}{\pi T_s}} \arcsin \left( \sqrt{\frac{\pi T_s }{2 J}}
  \rho \right) \right] + t^2} \, .
\end{equation}
This shape is plotted in figure~\ref{f:shape-string-xt}.  We see that
the solution depends on Minkowski time~$t$, whereas it is stationary
with respect to Rindler time~$\eta$. As time increases, we see that
(as expected) all points approach the light-cone~$x = \pm t$.  We also
see that the string configuration~\eqref{e:shapext} does not reduce to
a straight, unaccelerated string configuration at the initial
time~$t=0$, but that the string is already bent at this moment. This
is simply a consequence of the fact that we assume the acceleration to
be present at all times; a more realistic solution would start from a
straight string, with an acceleration only for~$t>0$, and would
exhibit more complicated time dependence, though with qualitatively
similar behaviour.

\subsection{Critical acceleration}

Let us now study the acceleration of the endpoints and see how it
depends on the spin. The constant~$C$ in~\eqref{e:stringshape} is the
acceleration of the midpoint of the string, and related to the
acceleration of the string endpoints. Evaluating~\eqref{e:stringshape}
at~$\pm L/2$ we find
\begin{equation}
\label{e:acceldef}
a^{-1} := \xi(\pm L/2) = \frac{C}{\kappa} \cosh\left[\frac{\kappa}{C}\sqrt{ \frac{\pi J}{2 T_s}}\right]\,,
\end{equation}
where we have defined the acceleration of the endpoints~$a$.  The
acceleration is limited from above by the minimum of the right-hand
side of~\eqref{e:acceldef} as a function of~$C$. For a generic value
of~$a$ there are two solutions, one of which is presumably
unstable~\cite{Friess:2006rk}. The maximum of~$a$ is achieved
for~$C/\kappa \approx 0.834 \sqrt{\pi J/2 T_s}$, resulting
in\footnote{This particular numerical value also plays a role for the
  computation of~\cite{Berenstein:2007tj} which involves static Wilson
  lines instead of rotating strings.}
\begin{equation}
\label{Acrit}
a \leq 0.529\sqrt{\frac{T_s}{J}}\, \equiv  a_{\text{max}}\,. 
\end{equation}
For this critical value of the acceleration the string world-sheet is
still smooth, and all points move within the light-cone of the string
endpoints. However, as we try to increase the acceleration beyond
$a_{\text{max}}$, the rigid U-shaped string solution ceases to exist. To
determine the precise time evolution of this over-accelerated meson, 
one would need to consider a more general, time-dependent
ansatz. However, a stability analysis of the critical string
configuration along the lines of~\cite{Friess:2006rk} makes it likely that
this configuration is unstable, leading to a final point of
dissociation given by two disconnected strings stretching all the
way to the horizon, and each carrying a fraction of the angular
momentum. 

From~\eqref{Acrit} we see that as expected, higher spin mesons are
less stable, and dissociate at a smaller value of the acceleration.

To estimate what is the value of the critical acceleration for
realistic mesons, we take the value of the string tension to be $T_s =
0.3\,\, \text{GeV}^2$.  In the holographic framework the description
of mesons in terms of large rotating strings is, strictly speaking,
only valid for values of angular momentum which are of the order $J
\sim \sqrt{\lambda}$, where $\lambda$ is the 't~Hooft coupling. It
would be interesting to extend our analysis to the (more realistic)
low spin meson sector, using a description in terms of probe-brane
fluctuations, but we will present this analysis elsewhere.  However,
if we naively take the expression~\eqref{Acrit} and evaluate it for
e.g.~$J=1,2$ and $10$, we find for the critical acceleration the
values $a_{\text{max}}=0.290,\, 0.205$ and $0.092\,\, \text{GeV}$
respectively. It would be interesting to see if there is an
experimental set up in which the critical acceleration could be
observed. In particular, the decelerations of nucleons in the initial
stage of a heavy ion collision is estimated to be of the order of up
to a GeV. Although our computation is done for mesons rather than baryons,
one may hope that a similar effect (with a similar order of magnitude
for the critical acceleration) should hold for baryons as well, and
hence may be relevant in describing the dynamics of the initial state of
the collision of heavy ions.

As mentioned before, the analysis in the previous section (the
ansatz~\eqref{ansz1}) and the value of the critical acceleration were
derived by assuming that the string accelerates in a direction
orthogonal to the rotation plane. The analysis of the equations of
motion describing acceleration under an arbitrary angle is complicated
and we will not attempt it here. However, in order to gain insight into
what is happening in this case, one can consider a simplified
configuration of a stretched string whose endpoints are accelerated in
the direction of their relative separation (with the same value of the
acceleration).  After a finite amount of lab time, the ``left'' endpoint
of the string will cross the Rindler horizon of the ``right'' endpoint,
and hence they will loose causal contact. The lab time of the crossing
is given by
\begin{equation}
t_{\text{cross}} = \frac{1}{2}\left( \frac{1}{L a^2} - L \right) \, , 
\end{equation}
where $L$ is the separation of the string endpoints in the lab frame.
There is, trivially, a critical value of the acceleration, $a_c$, for which the
crossing time is zero. If we assume that the separation of the string
endpoints originates from rotation, and we use~\eqref{JL}, the
critical acceleration is $a_c = \sqrt{\pi T_s/8 J}$. While the
dependence on the string tension is dictated by dimensional arguments,
it is interesting that this crude analysis still implies a dependence
on the angular momentum as in~\eqref{Acrit}. Hence, both longitudinal
and transverse accelerations lead to string melting behaviour.

We should note that in the computation above we have neglected
any effects of bremsstrahlung, since in the large-$N_c$ limit in which
our holographic computations are valid, the emission of strings
is suppressed by the string coupling~$g_s\sim 1/N_c$.

Finally, our computation was done in the simplifying limit of
vanishing constituent quark mass, although holographic models allow
for a description of mesons with non-zero constituent quark masses.
It would be interesting to extend our analysis to these more
realistic string configurations, and in particular, investigate
whether the inclusion of quark masses still leads to a universality of
the critical acceleration~\eqref{Acrit}, independent of the
supergravity background. It would also be interesting to analyse the
combined effect of a Rindler horizon in the four-dimensional
space-time and a holographic horizon in the radial direction. We will
leave this for future work.

\medskip

\section{Velocity dependent dissociation length}

In this section we shift our point of view from the one presented in
the previous sections, and use Rindler space (in the spirit
of~\cite{Berenstein:2007tj}) as a simplified gravitational background,
which will allow us to compute the velocity dependence of
the dissociation length, as obtained in full-fledged holographic models
in~\cite{Peeters:2006iu,Liu:2006nn,Chernicoff:2006hi}. The direction $\xi$ is now
interpreted as the holographic, fifth direction, and the acceleration
in this direction is caused by the gravitational curvature in the
holographic direction, rather than a genuine four-dimensional
acceleration.  We generalise the ansatz~\eqref{ansz1} to
\begin{multline}
\label{ansz2}
\eta = \eta(\tau)\,,\quad \xi=\xi(\sigma)\,,\quad\rho=\rho(\sigma)\,,\\[1ex]
\phi = \omega\tau\,,\quad y=v\eta(\tau)\,.
\end{multline}
where $(t, \rho, \phi,y)$ are directions on our 4-dimensional world,
and $\xi$ is the holographic direction. Note that the acceleration
(i.e.~finite temperature) breaks Lorentz invariance, which makes the
boosted solution inequivalent to the unboosted one.  Starting from the
ansatz~\eqref{ansz2}, the computation of the previous section is
trivially extended to yield the string profile
\begin{equation}
\label{e:stringshape2}
\xi(\sigma) = \frac{\sqrt{C^2 + v^2}}{\kappa} \cosh\left[ \frac{\kappa}{C} \sqrt{\frac{2 J}{\pi T_s}} \arcsin \left( \sqrt{\frac{\pi T_s}{2 J}} \sigma \right) \right]\, ,
\end{equation}
while the relation between the angular momentum $J$ and the length~$L$ of
the string in the $\rho$ direction is still given by~\eqref{JL}.

The endpoints of the string are now forced to sit on the flavour
D-brane, which is located at a fixed position~$\xi=\xi_0$. This
position is generically a function of the temperature of the
background and of the constituent quark mass. On this flavour brane,
the local speed of light is given by~$c_{\xi_0} = \kappa\xi_0$.  The constant
$C$ is again related to the position of the midpoint of the
string. 

\begin{figure}[t]
\vspace{4ex}
\includegraphics[width=.7\columnwidth]{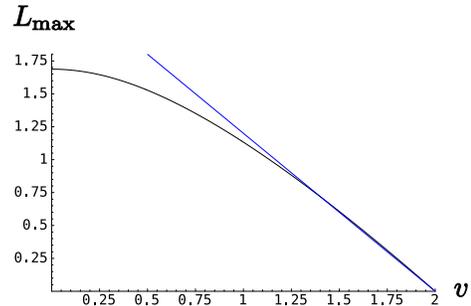}
\caption{The numerical solution for the dissociation length as a
  function of velocity, and the analytic large-$v$
  approximation.\label{small-big}\vspace{-1ex}}
\end{figure}

In contrast to the situation in the previous section, where we
minimised~$\xi(L/2)$ as a function of~$C$ for fixed~$J$, we thus now
want to find the value of~$C$ which maximises~$L$ for
fixed~$\xi_0$. An expression for~$L$ is obtained by
setting~$\xi(L/2)=\xi_0$ in the $v\not=0$ analogue of~\eqref{e:xiLC} and solving
for~$L$,
\begin{equation}
\label{e:acceldef2}
L = \frac{4C}{\kappa\pi} \arccosh\left[\frac{\kappa\xi_0}{\sqrt{C^2+v^2}}\right]\,.
\end{equation}
Unfortunately, an analytic solution for the maximum of~$L$ for all
values of the parameters~$v$ and $\xi$ seems out of reach. However,
the regime in which~$v \rightarrow \kappa\xi_0$ can be treated
analytically. This regime also corresponds to~$C\rightarrow 0$, and in
this limit the value of~$C$ which maximises~$L$ is
\begin{equation}
C = \sqrt{\frac{2}{3\,\kappa\xi_0}} v (\kappa^2\xi_0^2 - v^2)^{1/4}
\sqrt{\arccosh \frac{\kappa\xi_0}{v}}\,.
\end{equation}
Substituting this back into the expression for~$L$ and expanding
for~$v\rightarrow \kappa\xi_0$ leads to
\begin{equation}
\label{e:Lmax}
L_{\text{max}} = \frac{8 \sqrt{2}}{3 \pi \kappa} (\kappa \xi_0 - v) = \frac{4 \sqrt{2}}{3 \pi^2 T} (c_{\xi_0} - v)  
\end{equation}
where $T$ is the temperature of the Rindler horizon and $c_{\xi_0}$ is
the (local) speed of light.  This result indeed fits the full
numerical solution for~$L_{\text{max}}$ (see figure~\ref{small-big}), although in contrast
to~\cite{Peeters:2006iu,Liu:2006nn,Chernicoff:2006hi} there is no
agreement of this analytic result with a numerical solution in the
small-$v$ regime. 

Moreover, the form of the expression is different from the one
obtained in~\cite{Peeters:2006iu,Liu:2006nn,Chernicoff:2006hi} and we
also see that the result does not exhibit Lorentz invariance. The
reason for this difference is simple: standard holographic backgrounds
at finite temperature all tend to asymptotically flat space, unlike
the Rindler background~\eqref{Rind} which by construction describes
only the near horizon geometry of the non-extremal brane.  Asymptotic
flatness is a desired property of holographic backgrounds at finite
temperature, since it corresponds to the restoration of Lorentz
invariance at energies which are much higher than the temperature. The
metric~\eqref{Rind} and hence the result for the maximal dissociation
length~\eqref{e:Lmax} by construction do not describe this regime, but
they do capture the behaviour of the system at energies which are
smaller than or equal to the temperature. In this sense we work in a
regime which is complementary to that of~\cite{Liu:2006nn}.

\vspace{-3ex}

\section*{Acknowledgements}
\vspace{-1ex}

We thank Angel Paredes, Mukund Rangamani and Larry Yaffe for
discussions and especially Ofer Aharony and Cobi Sonnenschein for
comments on the manuscript and very useful correspondence. The work of
KP was supported by VIDI grant 016.069.313 from the Dutch Organization
for Scientific Research (NWO).


\end{document}